\documentclass[conference]{IEEEtran}
\IEEEoverridecommandlockouts
\usepackage{tcolorbox}
\usepackage[T1]{fontenc}
\usepackage{subcaption}
\usepackage{booktabs}
\usepackage{pgfplots}
\usepgfplotslibrary{statistics}
\pgfplotsset{compat=1.15}
\usepackage[hyphens]{url}
\usepackage{cite}
\usepackage{calc}  
\usepackage{enumitem} 
\usepackage{amsmath,amssymb,amsfonts}
\usepackage{algorithmic}
\usepackage{graphicx}
\usepackage{textcomp}
\usepackage{xcolor}
\usepackage{tcolorbox}
\usepackage{booktabs}
\usepackage[hidelinks]{hyperref}
\def\BibTeX{{\rm B\kern-.05em{\sc i\kern-.025em b}\kern-.08em
    T\kern-.1667em\lower.7ex\hbox{E}\kern-.125emX}}
\usepackage{flushend} 

\usepgfplotslibrary{statistics}
\pgfplotsset{compat=1.15}

\newcommand{\RqOne}{\textbf{(RQ1):} \emph{How prevalent are seed files in software repositories?}}

\newcommand{\RqTwo}{\textbf{(RQ2):} \emph{What kinds of seed files are there?}}

\newcommand{\RqThree}{\textbf{(RQ3):} \emph{Are there relationships among repositories in seed families?}}

\newcommand{\RqFour}{\textbf{(RQ4):} \emph{What was the main driver of the changes for variants?}}

\newcommand{\RqFive}{\textbf{(RQ5):} \emph{How uniquely do variants evolve in seed families?}}

\newcommand{\RqSix}{\textbf{(RQ6):} \emph{How do developers consider changes for variants?}}

\begin{document}

\title{Same File, Different Changes: The Potential of Meta-Maintenance on GitHub}

\author{\IEEEauthorblockN{Hideaki Hata,\IEEEauthorrefmark{1}
Raula Gaikovina Kula,\IEEEauthorrefmark{1}
Takashi Ishio,\IEEEauthorrefmark{1}
Christoph Treude\IEEEauthorrefmark{2}}
\IEEEauthorblockA{\IEEEauthorrefmark{1}Nara Institute of Science and Technology\\
\{hata, raula-k, ishio\}@is.naist.jp}
\IEEEauthorblockA{\IEEEauthorrefmark{2}University of Adelaide\\
christoph.treude@adelaide.edu.au}
}

\maketitle

\begin{abstract}
Online collaboration platforms such as GitHub have provided software developers with the ability to easily reuse and share code between repositories. With clone-and-own and forking becoming prevalent, maintaining these shared files is important, especially for keeping the most up-to-date version of reused code. Different to related work, we propose the concept of meta-maintenance---i.e., tracking how the same files evolve in different repositories with the aim to provide useful maintenance opportunities to those files. We conduct an exploratory study by analyzing repositories from seven different programming languages to explore the potential of meta-maintenance. Our results indicate that a majority of active repositories on GitHub contains at least one file which is also present in another repository, and that a significant minority of these files are maintained differently in the different repositories which contain them. We manually analyzed a representative sample of shared files and their variants to understand which changes might be useful for meta-maintenance. Our findings support the potential of meta-maintenance and open up avenues for future work to capitalize on this potential.
\end{abstract}

\section{Introduction}

Clone-and-own is a quick way to create customized variants of a software project by copying an existing 
product and adapting it to a new set of requirements~\cite{7884632,Dubinsky:2013:ESC:2495256.2495759,Ishio:2017:SFS:3104188.3104222}.
%
Despite perceived benefits, such as simplicity, availability, and independence of developers~\cite{Dubinsky:2013:ESC:2495256.2495759},
clone-and-own has been criticized for leading to a large number of code clones~\cite{Roy:2009:CEC:1530898.1531101}, making it difficult to propagate changes such as bug fixes from one instance to another.
Clone-and-own also leads to problems related to awareness: developers do not know when and where their code is being cloned, they do not know the origin of the cloned code, and they have no means of staying aware of changes to other instances that might benefit their own instance of the cloned code.


Source code reuse across multiple software projects has been widely studied in code clone research.
It is reported that a large number of projects included copies of libraries~\cite{6080795,7476786,Lopes:2017:DMC:3152284.3133908}.
Gharehyazie et~al.~reported that most projects obtain files from outside rather than providing their files to other projects~\cite{Gharehyazie:2017:HCC:3104188.3104225}.
Forking, including traditional hard forking~\cite{Ray:2012:CSC:2393596.2393659} and recent fork-based development in GitHub~\cite{Gousios:2014:ESP:2568225.2568260,Gousios:2016:WPC:2884781.2884826}, is one type of source code reuse. Recent studies discussed several problems in forking, such as redundant development, lost contributions, and fragmented communities~\cite{Zhou:2018:IFF:3180155.3180205,Zhou:2019:FSI:3338906.3338918}.

Even with its large-scale code resources and a large amount of developers, Google is reported to address reuse properly with a monolithic source code management system~\cite{Potvin:2016:WGS:2963119.2854146,Sadowski:2018:LBS:3200906.3188720}.
Such a monolithic system has several advantages: unified versioning, extensive code sharing and reuse, simplified dependency management, 
large-scale refactoring, 
and so on~\cite{Potvin:2016:WGS:2963119.2854146}.
Based on the monolithic source code management, changes to core libraries are promptly and easily propagated through the dependency chain into the final products that rely on the libraries~\cite{Potvin:2016:WGS:2963119.2854146}.

Although introducing such a complete monolithic source code management system in the entire free/libre open source software (FLOSS) ecosystem is not practical, tracking source code across multiple projects could bring new insight into software maintenance.
If we can aggregate and compose useful changes from various repositories maintaining the same origins, sharing such composed changes can help software maintenance activities efficiently for multiple projects. We call this approach \textit{meta-maintenance}.

To investigate the feasibility of meta-maintenance, in this paper, we report on the results of an exploratory study of almost 28 million files that are shared by multiple GitHub repositories. We found that in more than 70\% of the repositories in our sample, there is at least one file which also exists in another repository. Most of these files have not been maintained, but there is a significant minority which has not only been maintained but often has received project-specific changes, such as bug fixes. 
We manually analyzed a representative sample of such files as part of a qualitative analysis, and found that files that are shared by a large number of repositories are often libraries (e.g., jQuery) while files that are shared by a smaller group of repositories tend to contain utility functionality (e.g., drivers). In the former case, the repositories which share the file tend to be unrelated while in the latter case, there is often a relationship between repositories (e.g., one repository relying on the code in another).

Investigating how files changed in different repositories revealed a number of different types of changes, including library updates (e.g., upgrading the jQuery library), commits taken from a known origin (e.g., the Linux kernel project), as well as project-specific changes (e.g., bug fixes). We argue that project-specific changes have the largest potential for meta-maintenance, and we conducted a survey in which we asked project maintainers what they thought of specific instances where meta-maintenance could be applied in their repositories, i.e., the maintainers were maintaining a file that had been changed in another repository, potentially incorporating interesting changes. This survey result provided further evidence for the potential of meta-maintenance and pointed out interesting areas for future work.


\section{Related Work}
Before outlining our methodology, we discuss extensive literature related to code clones, origin analysis, and forks.

\subsection{Cross-Project Code Clones}

Code clone detection is the most popular approach to analyze source code reuse activities. 
Since developers may modify a code fragment for their own purpose, various tools have been proposed to detect similar source code fragments~\cite{Roy2009}.
Kamiya et~al.~\cite{KamiyaTSE2002} proposed CCFinder that analyzes normalized token sequences. 
Jiang et~al.~\cite{Jiang2007} proposed DECKARD that compares a vector representation of an abstract syntax tree. 
Nguyen et~al.~\cite{NguyenFASE2009} proposed Exas that compares a vector representation  of a dependence graph.
Sasaki et~al.~\cite{SasakiMSR2010} proposed FCFinder that recognizes file-level clones using hash values of normalized source files.
Cordy et~al.~\cite{Cordy:2011:NCD:2057176.2057234} proposed NiCad that compares a pair of code blocks using a longest common subsequence algorithm.
Sajnani et~al.~\cite{SajnaniICSE2016} proposed SourcererCC that compares a pair of code blocks using Jaccard index of tokens.

Code clone detection tools revealed that software developers often copy source files from other projects.
Hemel et~al.~\cite{Hemel2012} analyzed vendor-specific versions of Linux kernel. 
Their analysis showed that each vendor created a variant of Linux kernel and customized many files in the variant.
Ossher et~al.~\cite{6080795} analyzed cloned files across repositories using a lightweight technique.
They reported that projects cloned files from related projects, libraries, and utilities. 
Koschke et~al.~\cite{KoschkeIWSC2016} also reported that a relatively large number of projects included copies of libraries.
Lopes et~al.~\cite{Lopes:2017:DMC:3152284.3133908} analyzed duplicated files in 4.5 million projects hosted on GitHub and reported that the projects have a large amount of file copies.
%
%
Gharehyazie et~al.~\cite{Gharehyazie:2017:HCC:3104188.3104225} analyzed cross-project code clones of 5,753 Java projects on GitHub.  They also analyzed timestamps of the clones and reported that developers often copy an entire library, and some projects serve as hubs (sources) of clones to other projects.
The analyzed source code is a snapshot at a certain point of time.
Our study extends the analysis by including additional programming languages and all the versions of projects.

Identified code reuses across projects can be used for several applications.
Dang et~al.~\cite{Dang2017} reported that Microsoft developers use code clone information to fix bugs in multiple products at once.
Rubin et~al.~\cite{RubinSPLC2013} reported that industrial developers extract reusable components as core assets from existing software products.
Bauer et~al.~\cite{Bauer2013} proposed to extract code clones across products as a candidate of a new library.
Ishihara et~al.~\cite{IshiharaWCRE2013} proposed a function-level clone detection to identify reusable functions in a number of projects.
%
Luo et~al.~\cite{LuoFSE2014} proposed a method to identify semantically equivalent basic blocks for code plagiarism detection.
Chen et~al.~\cite{ChenICSE2014} proposed a technique to detect clones of Android applications using similarity between control-flow graphs of methods.


Davies et~al.~\cite{DaviesMSR2011,DaviesESE2013} proposed a file signature to identify the origin of a jar file using classes and their methods in the file ignoring the details of code. 
Ishio et~al.~\cite{Ishio:2017:SFS:3104188.3104222} extended the analysis to automatically identify libraries copied in a product.
Similar to these approaches, this study starts the analysis from file-level clones. 


\subsection{Origin Analysis}

Software projects use source code repositories to manage the versions of source code.  
Although a repository tracks modified lines of code between two consecutive versions of a file, the feature is not always sufficient to represent a complicated change.
Godfrey et~al.~\cite{Godfrey2005} proposed origin analysis to identify merged and split functions between two versions of source code.
The method compares identifiers used in functions to identify original functions.
Steidl et~al.~\cite{SteidlMSR2014} proposed to detect source code move, copy, and merge in a source code repository.
The method identifies a similar file in a repository as a candidate of an original version. 
Kawamitsu et~al.~\cite{KawamitsuSCAM2014} proposed an extension of origin analysis across two source code repositories.
Their method identifies an original version of source code in a library's source code repository.
%
Spinellis~\cite{Spinellis2016} constructed a Git repository including the entire history of Unix versions.
The unified repository enables developers to investigate the change history of source files across projects.
%
German et~al.~\cite{GermanMSR2009} used CCFinder to detect code siblings reused across FreeBSD, OpenBSD and Linux kernels, 
and then investigated the source code repositories of the projects to identify the original project of a code sibling.
Krinke et~al.~\cite{KrinkeIWSC2010} proposed to distinguish copies from originals by comparing timestamps of code fragments recorded in source code repositories.
Krinke et~al.~\cite{KrinkeMSR2010} used the approach and visualized source code reuse among GNOME Desktop Suite projects.
In this study, we do not intend to identify origins but investigate files in clone sets in terms of whether there are useful changes in their histories that could benefit other repositories.

\subsection{Forks}

Software developers often fork repositories in order to propose source code changes to the original projects.
Stanciulescu et al.~\cite{Stanciulescu:2015:FIV:2881297.2881381} analyzed an OSS community and reported that forks contribute new features while developers may spend their effort on redundant development.
Ren et al.~\cite{8668023} proposed a machine learning model using change description, patch content, and issue tracker to identify redundant code changes in forks.
Zhou et al.~\cite{Zhou:2018:IFF:3180155.3180205} proposed a tool named Infox to automatically identify unique source code changes as new features in forks.
Different from those analyses, this study does not focus on forked repositories, and revealed that non-forked repositories also include their own changes to reused source code.
Zhou et al.~\cite{Zhou:2019:FSI:3338906.3338918} reported that better modularity and centralized management are associated with more contributions and a higher fraction of accepted pull requests from forks.
They also reported that the lower the pull request acceptance rate, the higher the chance of a project having hard forks.
Regarding hard forking,
Ray et~al. analyzed porting of an existing feature or bug fix across forked projects~\cite{Ray:2012:CSC:2393596.2393659}.
They reported that forking allows independent evolution but results in the significant cost of porting activity.

Propagating changes to others is also considered as a challenge in software product line engineering.
Str\"{u}ber et al.~\cite{Struber:2019:FTB:3336294.3336302} included the task in scenarios for evaluating techniques that support developers during the evolution of variant-rich systems.
To support such propagation of a bug fix, ReDeBug~\cite{Jang2012} and Clorifi~\cite{Li2015} have been proposed to efficiently identify source file clones to which a patch should be applied. 

\section{Meta-Maintenance: Terminology}
\label{sec:terminology}

The concept of meta-maintenance is based on the model of individual evolution of the same files (clones) in different repositories (not limited to forks),
which is inspired by Google's monolithic source code management system~\cite{Potvin:2016:WGS:2963119.2854146}, with the aim to maintain the overall ecosystem.
We now define the terminology needed to describe meta-maintenance at file level.

A \textbf{seed file} is a specific version of a file that is shared in multiple software repositories. In this paper, we focus on Git repositories, thereby considering specific Git blobs appearing in multiple repositories to be seed files.
A Git blob stores the content of a file with a specific version, and is identified with its SHA-1 hash of the content~\cite{Chacon:2014:PG:2695634}.
Let $f$ refer to a seed file for a repository $r$ so that $r(f)$ indicates the repository $r$ containing file $f$.
Software repositories that contain the same seed file are then referred to as belonging to the same \textbf{seed family} $SF_0$, where $SF_0 = \{r_{1}(f), r_{2}(f),...,r_n(f)\}$ for repositories $r_1, r_2$ to $r_n$.
%
After the time $T$ passed, the seed family $SF_0$ becomes $SF_T = \{r_{1}(f_x), r_{2}(f_y),...,r_n(f_z)\}$, where $f_x$, $f_y$, and $f_z$ represent updated files from the same seed file $f$, and $x$, $y$, and $z$ are the number of changes to the files. The file $f_0 = f$, and we call $f_i (i > 0)$ a \textbf{variant}.
We call $f_l$ and $f_m$ \textbf{duplicate} variants if their contents are the same even if $l$ and $m$ are different.  
Meta-maintenance involves the analysis of how variants in the same seed family evolve.


Forking is an explicit form of sharing seed files.
In this paper we do not focus on forks as seed families. Recent studies have tried supporting change propagation in forks with a centralized model, that is, accumulating changes upstream and distributing them downstream~\cite{Zhou:2018:IFF:3180155.3180205,8668023,Zhou:2019:FSI:3338906.3338918}.
As well as such explicit reuse relationships, meta-maintenance is conceptualized to be able to support implicit reuse relationships.
We intend to support repositories in seed families with a decentralized model, that is, useful changes are aggregated from individual repositories and are distributed to others.

\section{Preparations}

In this section, we present our research questions and data collection methodology. 

\subsection{Research Questions}

The main goal of the study is to explore the potential for meta-maintenance for contemporary software projects in GitHub.
Based on this goal, we constructed six research questions to guide our study.
We now present each of these questions, along with our motivation.

\noindent
\RqOne~

\noindent
\RqTwo~

\noindent
\RqThree~

\noindent
\RqFour~

\noindent
\RqFive~

\noindent
\RqSix~

The motivation of \textbf{RQ1} is to understand whether seed files and their seed families are common in the wild. Furthermore, we would like to quantitatively explore the distribution, maintenance statuses, and patterns of seed families.
\textbf{RQ2}, \textbf{RQ3} and \textbf{RQ4} require a deeper analysis of seed files and repositories, where we would like to understand the nature of the seed files and reasons behind their evolution.
The key motivation for \textbf{RQ2} is to identify kinds of seed files that are used in the software repositories.
Such insights would help identify the nature for why these seed files were reused in other repositories.
Then for \textbf{RQ3}, we would like to understand whether there is a connection among repositories in the same seed families.
The key motivation for \textbf{RQ4} is to determine the key drivers that influenced the changes that were made to the seed files in those repositories.
%
\textbf{RQ5} then investigates the seed files from an evolutionary and maintenance standpoint. We would like to quantitatively understand how developers are updating or maintaining variants after reusing seed files in their projects and how differently variants evolve.
%
Our aim for \textbf{RQ6} is to understand how developers react to changes in other variants. 

\subsection{Data Collection}

Here we describe our procedures for target repository preparation, Git blob extraction, and stratified sampling, for our data collection.

\paragraph{Repository preparation}
To pursue the feasibility of meta-maintenance, we collect a large amount of software development repositories that have been actively developed.
We follow the same procedure as in a previous study~\cite{Hata:2019:9ML:3339505.3339656} to identify candidate repositories. We target software development repositories on GitHub written in seven \textit{common} programming languages, that is, C, C++, Java, JavaScript, Python, PHP, and Ruby. These languages have been ranked consistently in the top 10 languages on GitHub from 2008 to 2018 (based on the number of repositories from 2008 to 2015~\cite{github2015blog}, the number of pull requests from 2014 to 2018~\cite{githut}, and top languages from 2014 to 2018 in the official report~\cite{github2018octoverse}).
Using the GHTorrent dataset\footnote{MySQL database dump 2019-02-01 from \url{http://ghtorrent.org/downloads.html}.}~\cite{Gousios:2013:GDT:2487085.2487132}, we identify active repositories for the seven languages with the following criteria~\cite{Hata:2019:9ML:3339505.3339656}: (i) having more than 500 commits in their entire history (the same threshold used in previous work~\cite{Aniche:2018:CSM:3238579.3238606}), and (ii) having at least 100 commits in the most active two years to remove long-term less active repositories and short-term projects that have not been maintained for long. We determine repositories' languages based on the GHTorrent information.
As mentioned in Section~\ref{sec:terminology}, we exclude forked repositories in this study. We remove repositories that had been recorded in GHTorrent as forks of other repositories.
Note that even though we exclude such explicit forks, which have been targeted extensively by related work~\cite{Zhou:2019:FSI:3338906.3338918}, there can be implicit forks in our dataset. This could be because repositories were forked outside of GitHub, for example. Although these implicit forks are not considered in research on fork-based development~\cite{Zhou:2018:IFF:3180155.3180205,8668023,Zhou:2019:FSI:3338906.3338918}, we keep these repositories in this study to investigate the feasibility of meta-maintenance.
With the above procedure, we obtained the candidate list of repositories for the seven languages as shown in Table~\ref{tab:projects}.
From the candidate list, some repositories were not available because they had been deleted or made private. Additionally we exclude repositories that have only one committer in order to remove self-learning repositories (online judge code, for example).
In total, we obtained more than 32,000 repositories, which is almost 90\% of the candidate repositories.

\begin{table}
\caption{Collected repositories}
\centering
\label{tab:projects}
\begin{tabular}{lrr}
\toprule
\textbf{language} & \textbf{\# candidates} & \textbf{\# obtained (\%)} \\
\midrule
C & 3,262 & 2,962 (91\%) \\
C++ & 4,219 & 3,824 (91\%) \\
Java & 5,911 & 5,427 (92\%) \\
JavaScript & 9,017 & 7,960 (88\%) \\
Python & 6,606 & 6,087 (92\%) \\
PHP & 3,877 & 3,388 (87\%) \\
Ruby & 2,639 & 2,359 (89\%) \\
\midrule
\textbf{sum} & \textbf{35,531} & \textbf{32,007 (90\%)} \\
\bottomrule
\end{tabular}
\end{table}

\paragraph{Git blob extraction}
From each repository, we extracted all existing blobs and their file names using \texttt{git rev-list --all --objects} and \texttt{git cat-file} commands.
Only the blobs of source files with the following file extensions are targeted in this study: \texttt{.c}, \texttt{.h} (C), 
\texttt{.cc}, \texttt{.cp}, \texttt{.cpp}, \texttt{.cx}, \texttt{.cxx}, \texttt{.c++}, 
\texttt{.hh}, \texttt{.hp}, \texttt{.hpp}, \texttt{.hxx}, 
\texttt{.h++} (C++), 
\texttt{.java} (Java), \texttt{.js} (JavaScript), \texttt{.py} (Python), \texttt{.php} (PHP), and \texttt{.rb} (Ruby).
Blobs appearing in multiple repositories are considered to be seed files. In total, we obtained 27,994,587 seed files in our dataset.

\begin{figure}[h]
  \centering
  \includegraphics[width=\linewidth]{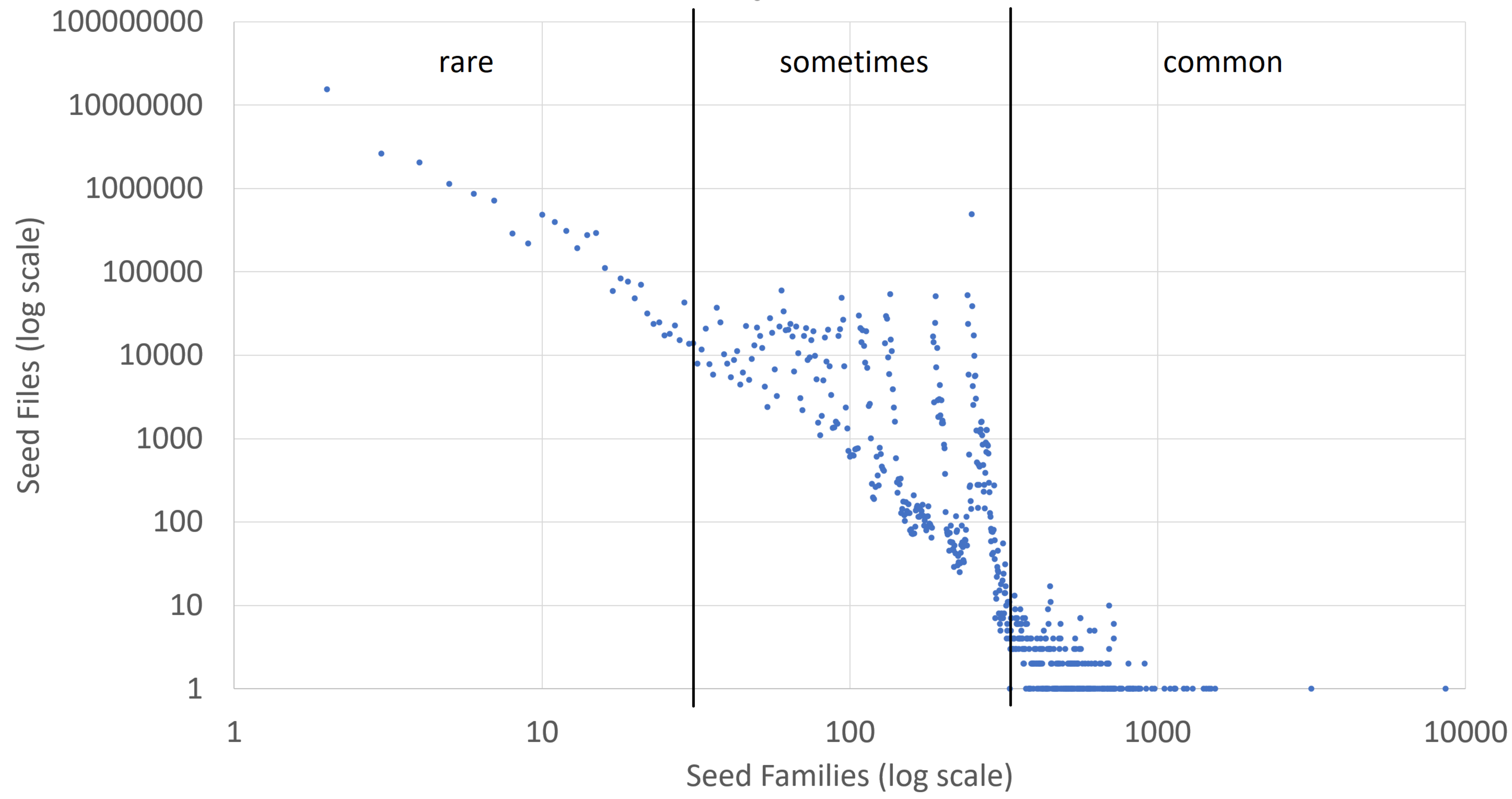}
  \caption{Plot of the number of seed families for each number of seed files (log-log scale).}
  \label{fig:distribution}
\end{figure}

\begin{table}
\centering
\caption{Construction of the stratified sample}
\label{tab:sampling}
\begin{tabular}{lrrr}
\toprule
 & \textbf{\# seed families} & \textbf{sample size} \\
\midrule
common & 662 & 243 \\
sometimes & 2,132,141 & 384 \\
rare & 25,861,784 & 384 \\
\midrule
\textbf{sum} & \textbf{27,994,587} & \textbf{1,011} \\
\bottomrule
\end{tabular}
\end{table}

\paragraph{Stratified sampling}
To understand these seed files and their characteristics towards answering our research questions, we conducted stratified sampling to enable us to gain insights into a variety of different scenarios. We hypothesize that the size of the corresponding seed families is a particularly important criterion for distinguishing different kinds of seed files. To understand the distribution of size of seed families across seed files, we plotted the distribution, cf.~Figure~\ref{fig:distribution}. Each dot represents a tuple of \textit{$<$number of seed families, size of seed family$>$}, e.g., the left-most dot indicates that there were more than 10 million seed families of size two. We then divided seed files into three categories: those that are part of big seed families (``common''), those are part of small seed families (``rare''), and an additional category in between (``sometimes''). Based on visual inspection of the plot in Figure~\ref{fig:distribution}, we set the thresholds at a family size of at least 331 for common, and at least 28 for sometimes. The left side of Figure 1 represents small seed families (rare) with less than 28 instances each and the right side represents large seed families (common) with more than 330 instances each.
Table~\ref{tab:sampling} shows the corresponding numbers. The stratum with a family size of at least 331 contains 662 seed families, the stratum with a family size of at least 28 (but no more than 330) contains more than 2 million seed families, and the stratum with a family size smaller than 28 contains more than 25 million seed families. The goal of our stratified sampling is to understand these different groups better. We randomly sampled a statistically representative number of seed families from each stratum, ensuring that our conclusions regarding ratios within each sample would generalize to the entire stratum with a confidence level of 95\% and a confidence interval of 5. The last column of Table~\ref{tab:sampling} shows the number of seed families in each sample, for a total sample size of 1,011. 

For the 1,011 seed families, the evolutionary period of variants within a seed family, measured from the earliest commit in seed files to the latest commit in variants, was a minimum of 0.4 years, a median of 5.5 years, and a maximum of 18 years.

\section{Method}

We describe our mixed-method procedure including a quantitative analysis, a qualitative analysis, and a survey.

\subsection{Quantitative Analysis}

To understand the prevalence of seed files in a large amount of software development repositories (\textbf{RQ1}), we conduct a quantitative analysis of our collected dataset and stratified sample in terms of existence, status, size, and modifications of seed files.
The history of a variant is examined based on the path of the seed file. Therefore, variants that have changed significantly in content can be tracked, but variants that have been renamed or moved are not tracked in this study.

To investigate how variants evolve in different projects (\textbf{RQ5}), 
we measure the degree of variant evolution in each family using two metrics: $Retention$ of similarities with seed files and $Uniqueness$ of differences with other variants.

\textbf{Retention: how variants are similar to seeds.}
For a pair of a variant $v$ and its seed file $s$, we measure their similarity as follows, similar to a previous study of clone-and-own~\cite{Ishio:2017:SFS:3104188.3104222}.
\begin{eqnarray*}
Retention_f(v, s) = \frac{| \{ \tau (v) \cap \tau (s) \} |}{|\tau (v)|} \\
\end{eqnarray*}
where $\tau(x)$ represents a set of trigrams of tokens in file $x$ ignoring comments and white spaces.
To recognize tokens and comments in source code, we employed lexical analyzers based on the Ripper library for Ruby and ANTLR4 for the other six languages.
%
We calculate an average retention value for all variants $V$ in a seed family as follows.
\begin{eqnarray*}
Retention(V, s) = \frac{\sum_{v \in V} Retention_f(v, s)}{|V|} \\
\end{eqnarray*}
A higher $Retention(V, s)$ means variants are similar to their seed files, that is, variants have not been changed largely.



\textbf{Uniqueness: how variants evolved differently.}
To measure the amount of project-specific changes for each variant, 
we calculate the uniqueness of content as follows.
\begin{eqnarray*}
Uniqueness(V, s) & = & \frac{\sum_{v \in V} u(v, \{ s \} \cup V \backslash \{ v \}  )}{|V|} \\
u(v, F) & = & \frac{| \{ t \in \tau (v) ~ | ~ \forall f \in F. t \notin \tau (f) \} |}{|\tau (v)|} \\
\end{eqnarray*}
The function $u(v, F)$ measures the amount of trigrams only in a variant $v$ in the family. A higher $Uniquness(V, s)$ value indicates that the variants are more different from one another.

\subsection{Qualitative analysis}

To understand the types of seed files (\textbf{RQ2}),
the relationships among repositories in seed families (\textbf{RQ3}), and
the characteristics of changes for seed files (\textbf{RQ4}),
we manually annotate seed files and seed families in our sample,
which is conducted in multiple iterations. In each iteration except for the last one, three authors independently annotate instances from each stratum using open coding. After each iteration, the authors discuss the codes that have emerged and how to distinguish between them. We repeat this process on new subsets from each stratum until we finalize the list of codes and achieve Kappa agreement of at least 0.7 among the three annotators. One author then annotates the rest of the data.
Since there can be a large number of variants in a seed family, we annotate up to five variants per seed family. If a seed family contains more than five variants, we conduct stratified sampling by selecting the three variants with the largest number of commits and the two variants with the lowest number of commits.

\subsection{Survey}

To understand the potential of meta-maintenance (\textbf{RQ6}), we conduct a survey for developer feedback.
To find meta-maintenance opportunities, we searched unique commits in seed families, that is, we identified commits that appear only in single repositories. To limit the search space, we target commits whose commit message contains the keyword `\textit{fix}', which is considered to be related to fixing bugs. There are 26, 81, and 135 seed families that contain at least one unique commit for rare, sometimes, and common samples, respectively. Again, to limit the search space, seed families with more than two unique commits are filtered out, which resulted in 11, 51, and 69 seed families for rare, sometimes, and common samples, respectively. We manually investigate these seed families. By checking the latest commits on GitHub, we found that some repositories applied the identified unique commits later by cherry-picking. We also found that some unique commits are not easily applicable to other repositories because of the large differences in the histories. 

Five cases were selected where the changes were neither too large nor too complex.
The survey was distributed in forums, issues, or emails of the corresponding projects.
As part of the questionnaire, we asked (a) how important the seed file was to the project, (b) what kind of maintenance activities were the developers interested in regarding the file, and (c) whether they would be interested in a meta-maintenance approach.
We received three responses and noticed that due to specificity of the files, only the core contributors of the projects responded in all cases.


\begin{figure*}[t!]
    \centering
    \begin{subfigure}[t]{\columnwidth}       
        \centering
			\includegraphics[width=\linewidth]{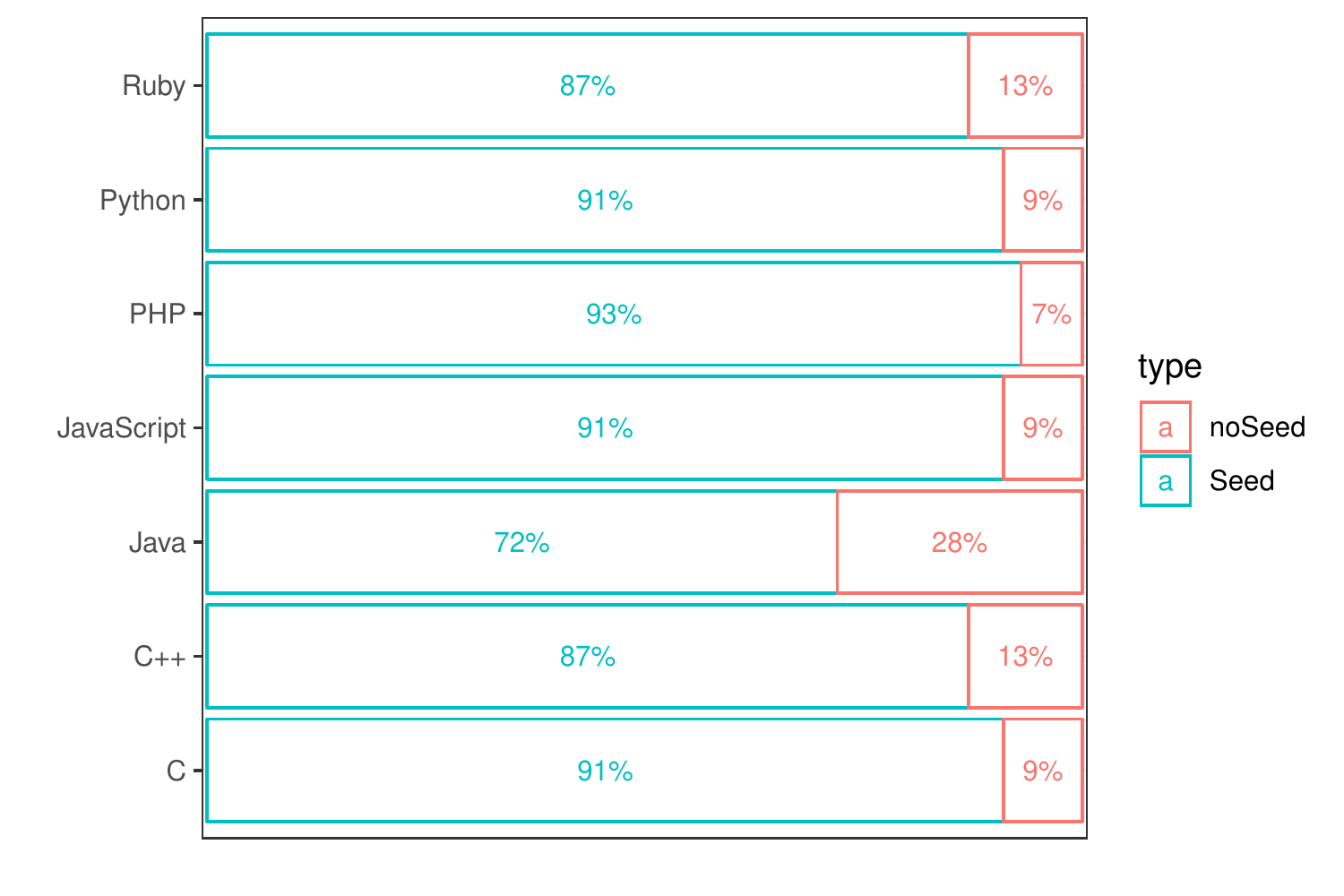}
		\caption{Percentage of repositories with and without seed files.}
	\label{fig:RQ1a}
    \end{subfigure}
    \begin{subfigure}[t]{\columnwidth}      
        \centering
			\includegraphics[width=\linewidth]{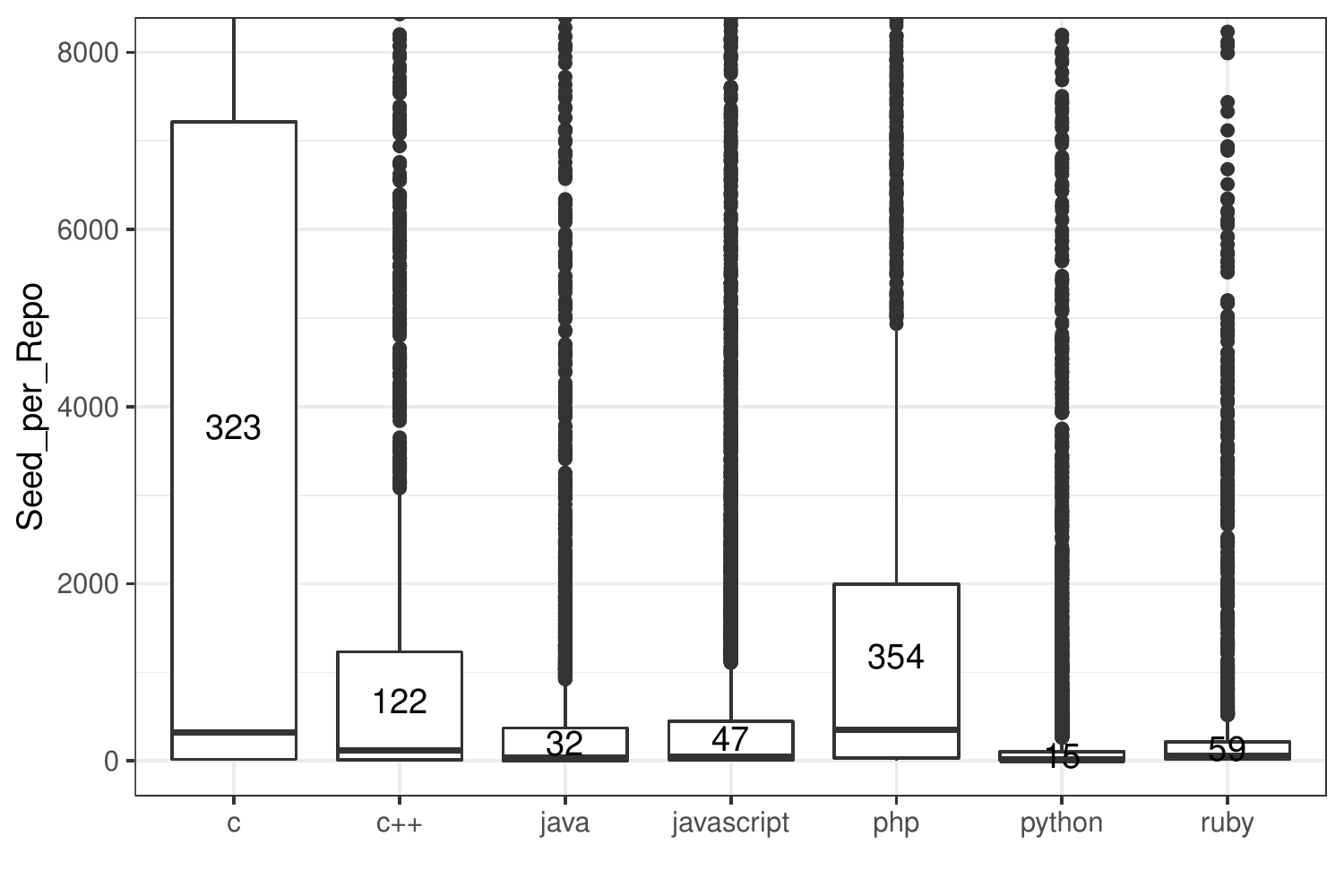}
		\caption{Distribution of the number of seed files per repository.}
	\label{fig:RQ1b}
    \end{subfigure}
    \caption{Prevalence of seeds from (a) seed file existence and (b) the number of seed files in a repository, per language.}
    \label{fig:RQ1}
\end{figure*}

\section{Findings}
\label{sec:find}

Here we present our findings for each research question.

\subsection{Prevalence of Seed Files (RQ1)}
\label{ssec:rq1}

\paragraph{Seed file existence}
Figure~\ref{fig:RQ1a} shows the percentages of repositories that have at least one seed file. Although the percentage is relatively low for Java (but still more than 70\%), more than 80\% of repositories contain seed files for the other languages. Especially for repositories written in C, JavaScript, PHP, and Python, more than 90\% of repositories contain seed files.
Figure~\ref{fig:RQ1b} presents the distributions of the number of different seed files per repository, for repositories that have at least one seed file. Median values are shown in the boxplot. We see that repositories contain dozens of seed files (at median), which implies there likely exist sets of seed files commonly shared by multiple repositories. For C and PHP, whose projects have more seed files, we see forked projects from specific products, which should share many seed files.

\begin{table}
\centering
\caption{Frequency of variant statuses in our sample}
\label{tab:status}
\begin{tabular}{lr@{}rr@{}rr@{}r}
\toprule
 & \multicolumn{2}{c}{\textbf{common}} & \multicolumn{2}{c}{\textbf{sometimes}} & \multicolumn{2}{c}{\textbf{rare}} \\
\midrule
dormant & 22,937 & (18\%) & 13,290 & (26\%) & 182 & (13\%) \\
inactive & 83,517 & (65\%) & 17,266 & (33\%) & 671 & (47\%) \\
unchanged & 15,435 & (12\%) & 1,278 & (2\%) & 52 & (4\%) \\
maintained & 7,530 & (6\%) & 20,181 & (39\%) & 518 & (36\%) \\
\midrule
\textbf{sum} & \textbf{129,419} & \textbf{(100\%)} & \textbf{52,015} & \textbf{(100\%)} & \textbf{1,423} & \textbf{(100\%)} \\
\bottomrule
\end{tabular}
\end{table}

\paragraph{Status of variants}
For the concept of meta-maintenance, we are interested in only repositories that are maintaining variants. We distinguish variants based on their statuses as follows.
\begin{description}
\item[dormant:] a variant is in a dormant or an unmaintained repository. Similar to previous studies~\cite{8101275,Coelho:2017:WMO:3106237.3106246}, we set one year as a threshold to consider dormant, that is, we consider a repository dormant or unmaintained if the repository does not have commits in 2018.
\item[inactive:] a variant does not exist in the main branch (usually \textit{master} in GitHub).
\item[unchanged:] the seed file has not been modified and exists in the latest commit of the main branch.
\item[maintained:] a variant had been changed and exists in the latest commit of the main branch.
\end{description}
Table~\ref{tab:status} shows the frequencies of the variant statuses in the three strata of our sample.
We see that the majority of variants are not \textit{maintained}, 94\% for common, 61\% for sometimes, and 64\% for rare.

\begin{table}
\centering
\caption{Frequency of seed family types in our sample}
\label{tab:type}
\begin{tabular}{lr@{}rr@{}rr@{}r}
\toprule
 & \multicolumn{2}{c}{\textbf{common}} & \multicolumn{2}{c}{\textbf{sometimes}} & \multicolumn{2}{c}{\textbf{rare}} \\
\midrule
not maintained & 0 & (0\%) & 48 & (13\%) & 240 & (63\%) \\
empty seed & 10 & (4\%) & 0 & (0\%) & 0 & (0\%)  \\
zero variance & 132 & (54\%) & 163 & (42\%) & 128 & (33\%)  \\
non-zero variance & 101 & (42\%) & 173 & (45\%) & 16 & (4\%)  \\
\midrule
\textbf{sum} & \textbf{243} & \textbf{(100\%)} & \textbf{384} & \textbf{(100\%)} & \textbf{384} & \textbf{(100\%)} \\
\bottomrule
\end{tabular}
\end{table}

\paragraph{Types of seed families}
Table~\ref{tab:type} summarizes the frequencies of seed family types as described below.

\begin{description}
\item[not maintained:] We observed that some projects recorded in GHTorrent point to the same repositories even though ID and/or project names are different. This happens when repositories re-registered in GitHub. We consider a seed family not maintained if there is no maintained variant in the seed family after removing identical repositories, that is, all variants in the seed family are either dormant, inactive, unchanged, or identical.
\item[empty seed:] We found some seed files do not have meaningful contents as source code, such as no line, only a blank line, or only comment lines. We manually identified those files. As seen in Table~\ref{tab:type}, such empty seed files appear only in the common stratum.
\item[zero variance:] There is only one \textit{maintained} variant or the same changes have been applied to all variants, that is, \textit{duplicate} variants. We found some variants had been cherry-picking commits from their origins with different frequencies. 
\item[non-zero variance:] Even after removing duplicate variants, there are multiple \textit{maintained} variants in a seed family.
\end{description}
For meta-maintenance, we are only interested in variants that have evolved independently. Therefore, only seed families of \textit{non-zero variance} are targeted in our study. In the seed families, duplicate variants are removed for the further analyses. Table~\ref{tab:sample} shows the number of seed families and the number of remaining variants in our filtered sample.
This sample is used to answer the following research questions.

\begin{tcolorbox}
\textbf{Summary}: We revealed that seed files are prevalent. In more than 70\% of the targeted 32,007 repositories in GitHub, there exists at least one seed file. Despite the large amount of seed files, most of them have not been maintained nor used in the latest snapshots. However, there exists some amount of potential variants for meta-maintenance for each stratum.
\end{tcolorbox}

\begin{table}
\centering
\caption{Target data in our sample}
\label{tab:sample}
\begin{tabular}{lrr}
\toprule
 & \textbf{\# seed families} & \textbf{\# variants} \\ 
\midrule
common & 101 & 897 \\ 
sometimes & 173 & 1,748 \\ 
rare & 16 & 47 \\ 
\midrule
\textbf{sum} & \textbf{290} & \textbf{2,692} \\ 
\bottomrule
\end{tabular}
\end{table}

\subsection{Types of Seed Files (RQ2)}
\label{ssec:rq2}

\begin{table}
\caption{Types of Seed Files}
\centering
\label{tab:rq2}
\begin{tabular}{lrrrr}
\toprule
          & \textbf{library} & \textbf{configuration} & \textbf{utility} & \textbf{other} \\
\midrule
common    & 91      & 5             & 2                     & 3     \\
sometimes & 5       & 1             & 157                   & 10    \\
rare      & 1       & 1             & 12                    & 2     \\
\bottomrule
\end{tabular}
\end{table}

We achieved Kappa agreement of 0.7 in our annotation.
Our analysis revealed the following four types of files, with Table~\ref{tab:rq2} showing the number of instances for each code in each stratum:

\begin{description}
\item[library:] Seed files which contain a library (a program that contains a collection of code used by applications) are particularly widespread in the common stratum where they account for 91\% of all seed files that we encountered during our annotation, indicating that most seed families in the common stratum are library users. A representative example is the jQuery library, e.g., in the modxcms-jp/evolution-jp repository.\footnote{\url{https://github.com/modxcms-jp/evolution-jp/blob/0dce7db4/manager/media/script/jquery/jquery.min.js}}
\item[utility functionality:] Seed files which contain utility functionality (a system software for controlling the operation of a computer, devices, etc.) were predominantly found in the sometimes stratum where they account for 91\% of the annotated seed files---coincidentally the same ratio as library files in the common stratum. Drivers, such as the driver for Avance Logic ALS300/ALS300+ soundcards in the masahir0y/linux repository\footnote{\url{https://github.com/masahir0y/linux/blob/dc4060a5dc25/sound/pci/als300.c}} are a representative example for this group of seed files.
\item[configuration:] Configuration files are much less common as seed files, with most of them appearing in the common strata. The config.php file of the contao/core-bundle\footnote{\url{https://github.com/contao/core-bundle/blob/1373ebc29/src/Resources/contao/config/config.php}} repository can serve as an example.
\item[other:] We used the code ``other'' mostly for header files or files containing version information, such as version.php in the gMagicScott/core.wordpress-mirror repository.\footnote{\url{https://github.com/gMagicScott/core.wordpress-mirror/blob/8548899126/wp-includes/version.php}}
\end{description}

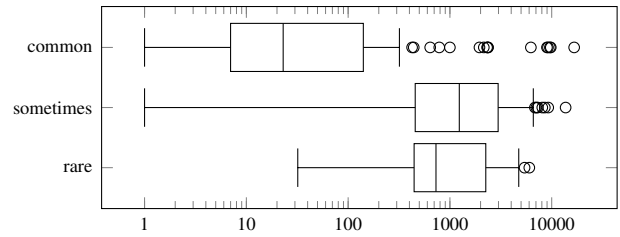
\begin{figure}
\centering
\begin{tikzpicture}
\tikzstyle{every node}=[font=\scriptsize]
\begin{axis}[y=8mm, try min ticks=2, xmode=log, xtick={1,10,100,1000,10000}, xticklabels={1,10,100,1000,10000}, ytick={1,2,3}, yticklabels={rare, sometimes, common}]

\addplot[black, mark=o, boxplot, color=black]
table[row sep=\\,y index=0] {
data\\
703\\ 35\\ 478\\ 756\\ 6024\\ 1850\\ 482\\ 5413\\ 4744\\ 32\\ 57\\ 1558\\ 481\\ 345\\ 1351\\ 3453\\
};

\addplot[black, mark=o, boxplot, color=black]
table[row sep=\\,y index=0] {
data\\
870\\ 991\\ 399\\ 1322\\ 246\\ 919\\ 984\\ 2872\\ 1212\\ 955\\ 1715\\ 1124\\ 4137\\ 1803\\ 102\\ 1944\\ 839\\ 778\\ 107\\ 3522\\ 146\\ 67\\ 220\\ 1236\\ 767\\ 1165\\ 3951\\ 1128\\ 182\\ 2537\\ 4651\\ 3470\\ 6837\\ 396\\ 1138\\ 139\\ 2521\\ 922\\ 478\\ 513\\ 3844\\ 56\\ 302\\ 85\\ 850\\ 1919\\ 270\\ 1943\\ 1826\\ 2669\\ 1123\\ 245\\ 127\\ 345\\ 2082\\ 2404\\ 690\\ 1781\\ 147\\ 456\\ 248\\ 2191\\ 3410\\ 2292\\ 5697\\ 3616\\ 2538\\ 223\\ 6345\\ 7066\\ 1366\\ 358\\ 6594\\ 691\\ 1249\\ 5738\\ 1471\\ 4590\\ 3051\\ 1735\\ 71\\ 1055\\ 563\\ 435\\ 733\\ 1629\\ 2914\\ 138\\ 3079\\ 139\\ 1537\\ 2116\\ 1024\\ 6081\\ 44\\ 1640\\ 2353\\ 1173\\ 140\\ 477\\ 241\\ 1746\\ 5528\\ 408\\ 1170\\ 873\\ 1842\\ 620\\ 1003\\ 3650\\ 2718\\ 339\\ 1348\\ 6175\\ 7285\\ 5798\\ 1872\\ 1073\\ 3684\\ 3075\\ 708\\ 3044\\ 1277\\ 984\\ 2621\\ 531\\ 878\\ 735\\ 13\\ 1\\ 3423\\ 4583\\ 2131\\ 2666\\ 8079\\ 60\\ 226\\ 147\\ 54\\ 5831\\ 211\\ 4655\\ 3012\\ 6244\\ 1522\\ 9240\\ 8557\\ 3174\\ 1579\\ 594\\ 862\\ 214\\ 1784\\ 443\\ 352\\ 152\\ 3000\\ 1067\\ 998\\ 1634\\ 2003\\ 6013\\ 1279\\ 2976\\ 3206\\ 623\\ 160\\ 13699\\ 961\\ 265\\ 6167\\ 4494\\ 4484\\ 
};

\addplot[black, mark=o, boxplot, color=black]
table[row sep=\\,y index=0] {
data\\
5\\ 1\\ 2363\\ 2\\ 48\\ 7\\ 23\\ 4\\ 205\\ 23\\ 2317\\ 209\\ 23\\ 999\\ 21\\ 319\\ 24\\ 2152\\ 16617\\ 3\\ 5\\ 79\\ 4\\ 4\\ 22\\ 4\\ 1\\ 23\\ 2\\ 13\\ 23\\ 31\\ 79\\ 786\\ 15\\ 35\\ 2363\\ 2\\ 23\\ 4\\ 39\\ 32\\ 18\\ 11\\ 23\\ 23\\ 94\\ 179\\ 23\\ 39\\ 32\\ 181\\ 1951\\ 4\\ 425\\ 167\\ 14\\ 7\\ 441\\ 9190\\ 23\\ 23\\ 23\\ 141\\ 4\\ 23\\ 23\\ 6\\ 3\\ 4\\ 18\\ 30\\ 60\\ 2\\ 13\\ 18\\ 75\\ 1\\ 5\\ 9789\\ 15\\ 23\\ 128\\ 6\\ 9597\\ 4\\ 4\\ 50\\ 23\\ 7\\ 5\\ 210\\ 205\\ 8981\\ 640\\ 64\\ 44\\ 178\\ 23\\ 11\\ 6240\\ 
};

\end{axis}
\end{tikzpicture}
\caption{LOC in seed files, per stratum.}
\label{fig:loc-per-strata}
\end{figure}

We also investigated typical file sizes in each stratum. Figure~\ref{fig:loc-per-strata} shows the corresponding distributions. In general, the number of lines of code for libraries, which are widespread in the common stratum, is smaller compared to the other strata, mostly due to presence of a large number of minified JavaScript libraries. The differences between sometimes and rare strata are negligible.

\begin{tcolorbox}
\textbf{Summary}: We revealed that seed files which are part of large seed families are often libraries, whereas seed files from medium-sized seed families tend to contain utility functionality. Files from large seed families tend to be smaller in terms of number of lines of code than files from smaller seed families.
\end{tcolorbox}

\subsection{Repositories in Seed Families (RQ3)}
\label{ssec:rq3}

As documented by previous work~\cite{munaiah2017curating}, GitHub hosts a wide variety of projects, the characterization of which is beyond the scope of this paper. Instead, to understand the potential of meta-maintenance, our focus is on the relationships among repositories in seed families, i.e., whether there is a connection among them. Such connections could stem from forking or copying as well as from the code of one repository using the code from another repository. 
We distinguish each repository in our sample into ``related'' or ``non-related''. 
In this annotation, we achieved perfect Kappa agreement (i.e., 1.0) among the three annotators.

\begin{description}
\item[related:] There is an explicit relationship among repositories, e.g., one is a fork of another, their names are similar or identical, or because one mentions the other prominently in its documentation. The most common example of such a relationship are the many Linux variants in our sample. For example, the driver file gl520sm.c is contained in the repositories masahir0y/linux\footnote{\url{https://github.com/masahir0y/linux}} and Whissi/linux-stable.\footnote{\url{https://github.com/Whissi/linux-stable}} The repositories can easily be connected based on their names, even though only the former contains the information that it is forked from torvalds/linux. In addition, the latter repository states in its description that it is a mirror of the Linux distribution hosted on git.kernel.org.\footnote{\url{https://git.kernel.org/pub/scm/linux/kernel/git/stable/linux.git}}
\item[non-related:] On the other hand, many seed families do not contain any evidence to suggest that there is a connection among the repositories, apart from using at least one common file. For example, the repositories MoveLab/tigatrapp-server\footnote{\url{https://github.com/MoveLab/tigatrapp-server}} and magda-io/magda\footnote{\url{https://github.com/magda-io/magda}} both contain Twitter's Bootstrap library via a bootstrap.js file. The former repository contains a server with which Tigatrapp (a Spanish citizen science project) apps communicate whereas the latter contains an open-source software platform designed to assist in all areas of the data ecosystem. Apart from both repositories using Bootstrap, there is no apparent connection among the repositories. Note that both repositories have made one change to their instance of bootstrap.js since the files were identical.
\end{description}

\begin{table}
\caption{Relationships among Seed Families}
\centering
\label{tab:rq3}
\begin{tabular}{lrr}
\toprule
          & \textbf{related} & \textbf{non-related} \\
\midrule
common    & 0       & 398         \\
sometimes & 624     & 24          \\
rare      & 40      & 5           \\
\bottomrule
\end{tabular}
\end{table}

Table~\ref{tab:rq3} shows that the ratio of related vs.~non-related differs significantly between the strata. While all repositories in our sample from the common stratum were not related, the vast majority from sometimes and rare were related.

\begin{tcolorbox}
\textbf{Summary}: Repositories which contain seed files that are part of large seed families tend to not be related. In many cases, they are different repositories using the same library. On the other hand, repositories which contain seed files which are part of smaller seed families tend to be related. 
\end{tcolorbox}

\subsection{Changes for Seed Files (RQ4)}
\label{ssec:rq4}

To understand how the repositories might be able to benefit from meta-maintenance, we investigated the commit histories of all files in our sample to characterize the set of changes that had been applied to them. 
Our annotation 
revealed four categories plus other, and we achieved Kappa agreement 
of 0.76.

\begin{table}
\caption{Change Types}
\centering
\label{tab:rq4}
\begin{tabular}{lrrrrr}
\toprule
          & \textbf{known} & \textbf{library} & \textbf{project-} & \textbf{tangled} & \textbf{other} \\
          & \textbf{origin} & \textbf{updates} & \textbf{specific} & \textbf{updates} &       \\
\midrule
common    & 9            & 202             & 96               & 15      & 76    \\
sometimes & 619          & 11              & 11               & 3       & 4     \\
rare      & 34           & 0               & 8                & 0       & 3     \\
\bottomrule
\end{tabular}
\end{table}

\begin{figure*}[t!]
    \centering
    \begin{subfigure}[t]{\columnwidth}       
        \centering
\includegraphics[width=\linewidth]{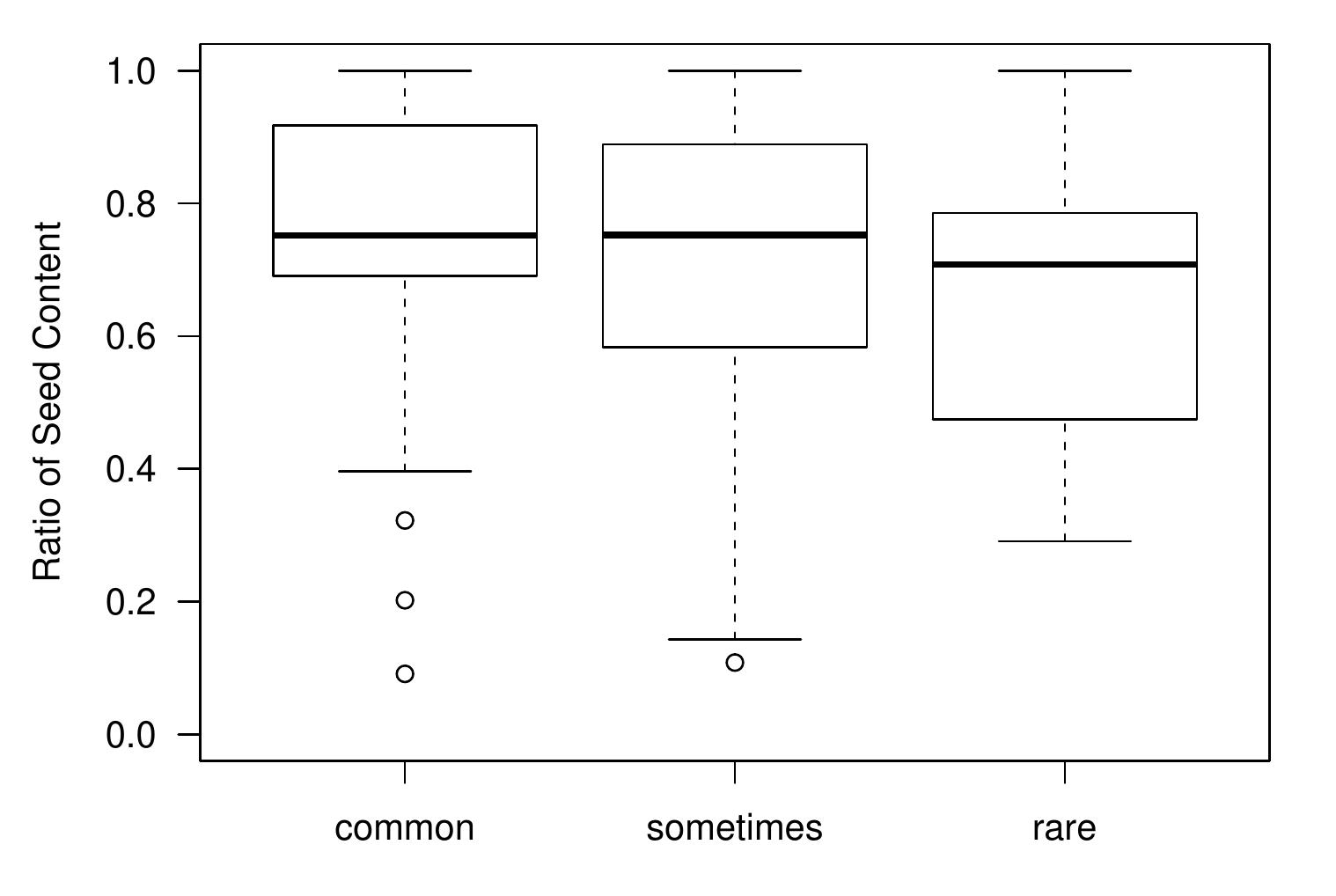}
\caption{Distribution of $Retention(V, s)$, per stratum.}
\label{fig:retention}
    \end{subfigure}
    \begin{subfigure}[t]{\columnwidth}      
        \centering
\includegraphics[width=\linewidth]{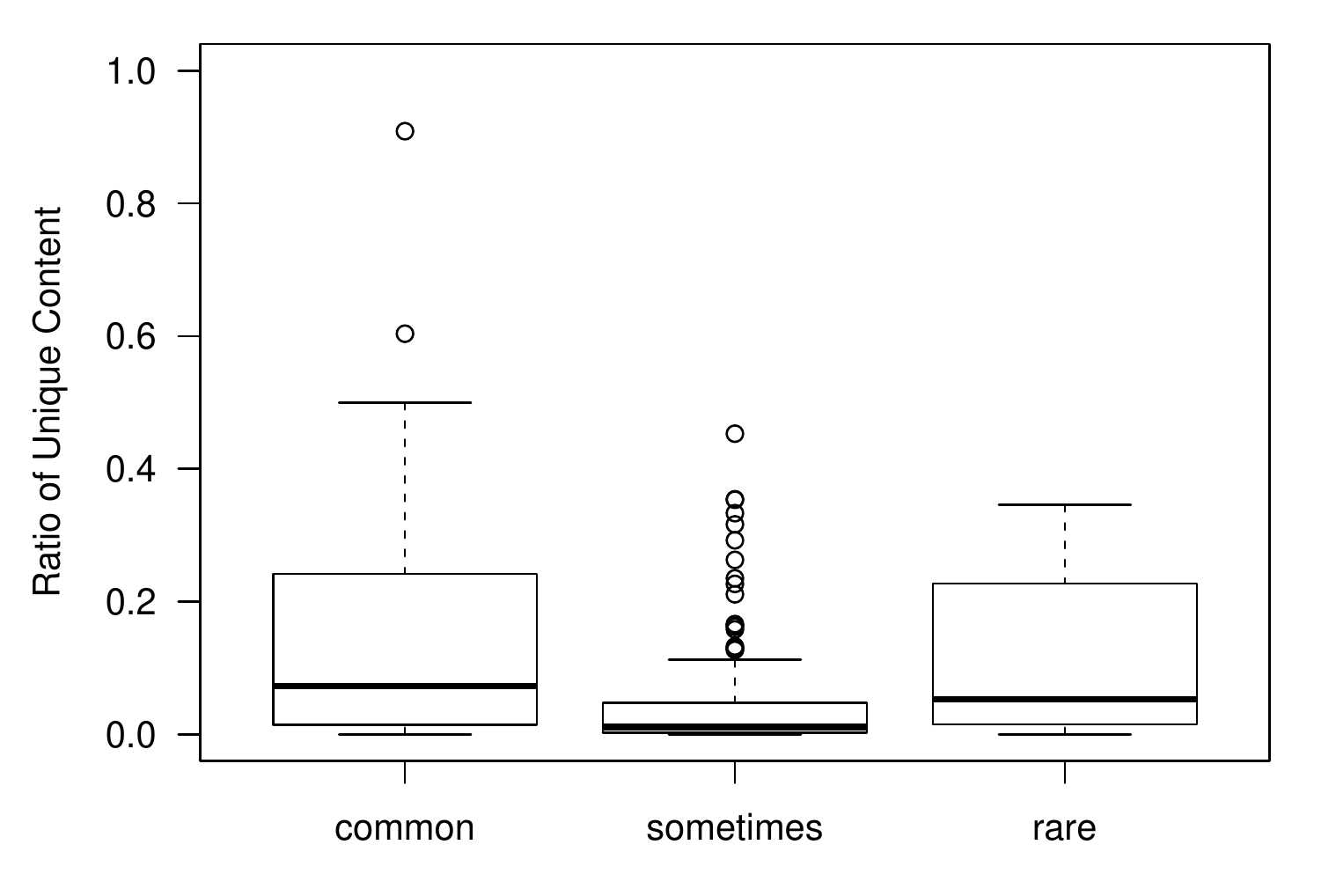}
\caption{Distribution of $Uniqueness(V, s)$, per stratum.}
\label{fig:uniqueness}
    \end{subfigure}
    \caption{Evolution of variants from (a) similarities with seed files and (b) differences with other variants.}
    \label{fig:RQ5}
\end{figure*}

\begin{description}
\item[reference to a known origin:] For many repositories in our sample, the origin is obvious---this applies in particular to the various Linux variants. In those cases, changes that have been applied to seed files are often the same commits that have been applied to the origin. We were often able to detect this directly from the commit meta information, e.g., when commit author and commit committer are different. For example, the process.c file\footnote{\url{https://github.com/Ziyann/omap/blob/7c07453808b/arch/powerpc/kernel/process.c}} in the Ziyann/omap repository attracted 104 commits since being identical across its seed family, many of which were committed by Linus Torvalds\footnote{\url{https://github.com/torvalds}} and authored by a contributor of the Ziyann/omap repository.
\item[library updates:] Library updates are the most common kind of change in the common stratum, see Table~\ref{tab:rq4}. An example is the previously mentioned jQuery library, e.g., in the modxcms-jp/evolution-jp\footnote{\url{https://github.com/modxcms-jp/evolution-jp}} repository. While there have been 18 commits to this file since it was identical across its seed family, the commit messages of these commits\footnote{\url{https://github.com/modxcms-jp/evolution-jp/commits/0dce7db475116f3a35206714e2721bf355f049c2/manager/media/script/jquery/jquery.min.js}} clearly show the pattern of library updates, e.g., ``Update jQuery 1.11.0 --$>$ 1.11.1'' and ``Update -- jQuery 3.2.1''.
\item[project-specific changes:] The most interesting case for meta-maintenance are project-specific changes which are not library updates or made in reference to a known origin. As Table~\ref{tab:rq4} shows, we found such files in all strata. An example is the jQuery library\footnote{\url{https://github.com/dmitrykuzmin/chat/blob/fa33c7e/webapp/src/main/webapp/lib/jquery-2.1.4.min.js}} in the dmitrykuzmin/chat repository which has one new commit since being identical across its seed family with the commit message ``Trying to fix IE issues.''. We argue that such fix attempts might be relevant to other repositories containing the same file. We found project-specific changes in at least one quarter of seed families in the common and rare strata. 
\item[tangled updates:] In case the commit history contained changes that could not easily be localized to the file under investigation, we applied the code ``tangled updates''. In these cases, meta-maintenance on the basis of files is unlikely going to be successful since changes affected many files---often hundreds if not thousands. Tangled updates were not widespread in our sample.
\item[other:] For change histories which did not fit any of the previous categories, we applied the code ``other''. An example are the commits to bootstrap.js\footnote{\url{https://github.com/sparc-request/sparc-request/blob/0eb424aba/app/assets/javascripts/bootstrap.js}} in the sparc-request/sparc-request repository. All of the four commits made since the file was identical across seed families indicate updates to the copyright information, most of them simply changing the year. This is unlikely to be useful for maintenance.
\end{description}

\begin{tcolorbox}
\textbf{Summary}: While some variants were updated in references to a known origin or as part of regular library updates, in particular seed files that are part of large (common stratum) or small (rare stratum) seed families also contain a significant number of project-specific changes which could be useful for meta-maintenance.
\end{tcolorbox}

\subsection{Uniqueness of Variants (RQ5)}
\label{ssec:rq5}

Figure~\ref{fig:retention} shows the distributions of the $Retention$ values for each stratum.
Their medians are 0.75, 0.75, and 0.71, for common, sometimes, and rare, respectively.
Most variants include some changes from the seed file.
While the rare families seem to include more changes, 
the differences among the strata are not statistically significant.
The Wilcoxon Rank Sum Test results in $p=0.267$ for the rare-sometimes pair, $p=0.065$ for the sometimes-common pair, and $p=0.097$ for the rare-common pair.

Figure~\ref{fig:uniqueness} shows the distributions of the $Uniqueness$ values for each stratum.
The medians are 0.09, 0.01, and 0.07, for common, sometimes, and rare, respectively.
The Wilcoxon Rank Sum Test shows that sometimes families have significantly lower uniqueness than rare and common families ($p=0.005$ for the rare-sometimes pair and $p<0.001$ for the sometimes-common pair).
The difference between rare and common families is not significant ($p=0.674$).
%
The variants in the common and rare families include more unique changes.
Those unique changes are potentially useful for other projects in the same family.
For the sometimes families, variants tend to evolve by following the changes in their seed files but do not include much unique content.
This result confirms the findings from our manual investigation (Section~\ref{ssec:rq4}).



\begin{tcolorbox}
\textbf{Summary}: Variants in the common and rare families contain project-specific changes in terms of unique tokens, while variants in the sometimes families tend to share common tokens.
Although our analysis provides the basis for understanding how uniquely variants evolve, further research is needed to understand seed files in terms of evolution and maintenance.
\end{tcolorbox}

\subsection{Developer Feedback on Meta-Maintenance (RQ6)}
\label{ssec:rq6}

We present three cases where we received responses from developers.

\textbf{Case Study 1: variants of \texttt{jQuery.js} library.}
\textsf{jQuery} is a widely used JavaScript library. Although the latest version in March 2020 is 3.4.1, some projects had been using older 1.x versions because of project-specific reasons.
\textsf{Joomla}\footnote{\url{https://github.com/joomla/joomla-cms}} is one such project which had applied bug fixing related to 
%
security issues.\footnote{
\url{https://github.com/joomla/joomla-cms/commits/a81ada410a5bf6b700a79d432fc5926146ac9f94/media/jui/js/jquery.js}}

The survey was completed by an experienced core developer of \textsf{Joomla}, who in the comments thanked us for bringing up this topic.
When asked about specific modifications to the same file in another repository, 
the respondent revealed that they are ``\textit{Mainly interesting for security reasons}'' and gave us the following detailed comment.
%
\begin{quote}
    \textit{For 3rd party components, we use tools like npm or composer to update them to the latest version. Only if we have to support a particular version we maintain the files our self and only for security fixes.}
\end{quote}

This answer reveals an interesting use case for the concept of meta-maintenance: the adoption of security-related changes that have already been applied elsewhere. As a previous study reported, vulnerability issues are not appropriately addressed in many software development repositories~\cite{Kula:2018:DUL:3188697.3188710}.
Aggregating appropriate fixes and distributing to individual repositories may support such cases.

\textbf{Case Study 2: variants of \texttt{abc\_object.cc} in Blender forks.}
\textsf{Blender} is an open source project for 3D computer graphics. In the rare families of our sample, there are two forks: \textsf{Bforartists}\footnote{\url{https://github.com/Bforartists/Bforartists}} and \textsf{blender278}.\footnote{\url{https://github.com/tangent-animation/blender278}}
The former is an active project merging all the recent commits from the origin without project-specific changes, while the latter does not have commits in the last half year but has project-specific changes from before. Both projects maintained the seed file \texttt{abc\_object.cc} at least in 2018. After forking from the same content, the former had been changed by 31 commits and the latter had been changed by 13 commits. 

The survey was completed by an experienced core developer of \textsf{Bforartists}, who claimed that the file is of high priority (5/5 on the Likert scale). 
When asked about modifications to the same file in another repository, 
the response was as follows.
\begin{quote}
    \textit{Relevant is the Blender source code, since we regularly merge the newest changes from Blender source code. We are not interested in another fork with probably outdated code, or code that conflicts with our changes.}
    \textit{... maintaining code from others is always problematic. It's hard enough to keep our own changes working. We are a fork of Blender, not a fork of another fork..}
\end{quote}

These comments point out important future work to make meta-maintenance a reality. 
To maintain seed files, some forked projects prefer to keep following changes in the original repository and do not consider applying changes from other repositories. This is a reasonable approach when centralized fork models work. We can easily identify repositories that are only following the original repository---meta-maintenance appears to be less relevant for them. In addition, we can consider aggregating project-specific changes and send them to the original repositories.


\textbf{Case Study 3: variants of \texttt{zlib} library.}
\textsf{zlib} is a library for data compression, which is widely used in many projects.
We observed a local change which fixes typos in one repository.\footnote{\url{https://github.com/radareorg/radare2/commit/bc3425e73d294cbded877b66f0d60183edb5dd2e}} The survey was completed by the author of \textsf{zlib}. Although the developer was not interested in the suggested typo fix, he described the potential of applying commits from other repositories as:

\begin{quote}
\textit{They may have performance or other improvements.}
\end{quote}

\begin{tcolorbox}
\textbf{Summary}:
We learned from developers that supporting changes related to security and performance, among others, could be a promising use case for meta-maintenance and is desired by developers.
Responses also point out future work to further understand the nature of relationships between repositories.

\end{tcolorbox}

\section{Discussion}
Based on our results, we now summarize the open challenges and barriers that need to be addressed before meta-maintenance can be fully realized.
Then we discuss the limitations of our study.

\subsection{Challenges to Meta-Maintenance}
This paper 
establishes the state-of-the-practice for investigation of what type of meta maintenance is useful and under which conditions.
To fully realize the potential of meta-maintenance, further research is required in the follow areas:

\begin{itemize}
    \item \textit{In-depth investigation of clone-and-own relationships in sets of seed files}.
    This study only focuses on single files as seeds. However, it is natural for software development projects to reuse a set of files.
    We need to develop techniques to identify seed families by considering a set of files, similar to a previous study~\cite{Ishio:2017:SFS:3104188.3104222}.
    During our manual investigation in \textbf{RQ6}, we observed that some seed families consist of sub-groups of repositories. Repositories in a group evolve similarly, and this results in large differences in the histories across different sub-groups. Measuring the similarities of histories (similar to the analysis in \textbf{RQ5}) is a promising and challenging area for future work.
    \item \textit{Developing a global source code tracking system}, to understand relationships of repositories and the histories of seed files. This could be similar to Google's monolithic SCM system~\cite{Potvin:2016:WGS:2963119.2854146,Sadowski:2018:LBS:3200906.3188720}, but in the ecosystem of open source software projects. The findings from \textbf{RQ1} reveal that many repositories could be connected within this global system.
    \item \textit{Techniques to identify useful changes between variants}.
    There exist related techniques for fork-based development~\cite{Zhou:2018:IFF:3180155.3180205,Ren:2019:APP:3338906.3342488} and software product lines~\cite{Pfofe:2016:SSV:2934466.2962726}.
    Extending existing characterization studies, e.g., repeated bug fixes~\cite{8094441} to clone-and-own instances may provide further insights to develop such techniques.
    As learned from the findings in \textbf{RQ6}, changes related to security and performance are promising types to be identified as useful changes.
    \item \textit{Tool support for meta-maintenance}. 
     Tools to help developers find specific changes and maintain code automatically could be practically useful. Such tools could be based on push and pull models to aggregate and distribute useful changes to the ecosystem.
\end{itemize}

\subsection{Threats to Validity}

Threats to \textit{construct validity} exist in our data collection procedure.
Our criteria for selecting repositories may have ignored recent active projects and projects worked on by a single developer. Although we categorized repositories by programming language based on GHTorrent information, the information can be inaccurate. Further exploration to develop better criteria is needed.
%
Threats to  \textit{external validity} exist in our repository preparation. Although we analyzed a large amount of repositories on GitHub, we cannot generalize our findings to industrial projects nor FLOSS in general; some FLOSS repositories are hosted outside of GitHub, e.g., on GitLab or private servers. In addition, empirical studies are needed for other programming languages.
To mitigate threats to \textit{reliability}, we prepare an online appendix of our studied dataset with associated information (see Section~\ref{ssec:appendix}).

\section{Conclusion}
\label{sec:cfw}
To explore the potential of meta-maintenance, we conducted an exploratory study with
(i) a quantitative analysis of 27,994,587 seed files from 32,007 Git repositories to establish the prevalence of seed files, the extent to which seeds evolve, and the uniqueness of seeds;
(ii) a qualitative analysis of a stratified sample of 1,011 seed files to determine the kinds of seeds, the relationships among seed families, and main drivers for changes in the variants; and
(iii) a survey for developer feedback.

Our work shows the potential of meta-maintenance with an extensive type of changes identified other than simple forking.
Based on this work which has established the prevalence of seeds in GitHub projects, their multiple categories of seed variants, uniqueness and practical useful potential of meta-maintenance, there are many open avenues and challenges for future work: understanding how to manage all seed variants in seed families, further studies of what are useful changes, and tool support to extract specific needs of a seed family to query other repositories, to name a few.

\section{Data Availability}
\label{ssec:appendix}

Our online appendix contains the list of the studied 32,007 repositories on GitHub, the list of the targeted 401,610,677 files, the results of the qualitative and quantitative analyses, and survey material.
The appendix is available at \url{https://github.com/NAIST-SE/MetaMaintenancePotential}.

\section*{Acknowledgment}

This work has been supported by JSPS KAKENHI Grant Numbers JP16H05857, JP18KT0013, JP18H04094, JP20K19774, and JP20H05706.


\flushend
\end{document}